\documentclass[12pt]{article}
\usepackage[utf8]{inputenc}
\usepackage[margin=0.75in]{geometry}
\usepackage{amsmath}
\usepackage{amssymb}
\usepackage{bm}
\usepackage[round,authoryear,sort]{natbib}
\usepackage[colorlinks=true,allcolors=blue]{hyperref}
\usepackage{lmodern}

\usepackage[T1]{fontenc}
\usepackage{tikz}
\usepackage{float}
\usepackage{xcolor}
\usepackage{listings}

\definecolor{mygray}{rgb}{0.95,0.95,0.95}
\definecolor{mygreen}{rgb}{0,0.4,0}
\definecolor{myblue}{rgb}{0,0,0.6}

\lstdefinestyle{R}{
    language=R,
    backgroundcolor=\color{mygray},
    commentstyle=\color{mygreen},
    keywordstyle=\color{myblue},
    basicstyle=\footnotesize\ttfamily,
    numbers=left,
    stepnumber=1,
    alsoletter={_},
    numberstyle=\footnotesize\ttfamily,
    showstringspaces=false,
    otherkeywords = {!,!=,~,*,\&,\%/\%,\%*\%,\%\%,<-,<<-},
    morekeywords={plot_observations,plot_forecasts,summary_stats,plot_score_by_leadtime,make_evaluation_subset,evaluate_marginal_distribution,rank_histogram_list,event_detection_table,roc_curve_list,reliability_diagram_list,brier_score_list,contingency_table_list,evaluate_joint_distribution},
    deletekeywords={data,all,grid,by,range,window}
}

\lstdefinestyle{output}{
    language=R,
    backgroundcolor=\color{mygray},
    basicstyle=\footnotesize\ttfamily
}

\title{evalprob4cast: An R-package for evaluation of ensembles\\as probabilistic forecasts or event forecasts}
\author{Mathias Blicher Bjerregård, Jethro Browell, John Zack, Jan Kloppenborg Møller,\\ Henrik Madsen, Gregor Giebel and Corinna Möhrlen}

\begin{document}

\newcommand{\datA}{\texttt{simplewind}}
\newcommand{\datB}{\texttt{oneyearwind}}
\newcommand{\Rpack}{\texttt{evalprob4cast}}
\setlength{\parindent}{0pt}

\maketitle

\begin{abstract}
    For any forecasting application, evaluation of forecasts is an important task. For example, in the field of renewable energy sources there is high variability and uncertainty of power production, which makes forecasting and the evaluation hereof crucial both for power trading and power grid balancing. In particular, probabilistic forecasts represented by ensembles are popular due to their ability to cover the full range of scenarios that can occur, thus enabling forecast users to make more informed decisions than what would be possible with simple deterministic forecasts. The selection of open source software that supports evaluation of ensemble forecasts, and especially event detection, is currently limited. As a solution, \Rpack\text{ }is a new R-package for probabilistic forecast evaluation that aims to provide its users with all the tools needed for the assessment of ensemble forecasts, in the form of metrics and visualization methods. Both univariate and multivariate probabilistic forecasts as well as event detection are covered. Furthermore, it offers a user-friendly design where all of the evaluation methods can be applied in a fast and easy way, as long as the input data is organized in accordance with the format defined by the package. While its development is motivated by forecasting of renewables, the package can be used for any application with ensemble forecasts.
\end{abstract}
\section{Introduction}

This paper documents the content and use of a new R-package for probabilistic forecast evaluation, \Rpack, available on GitHub: \hyperlink{https://github.com/jbrowell/evalprob4cast}{https://github.com/jbrowell/evalprob4cast}. It serves as a self-contained open source implementation of the \textit{IEA Wind Recommended Practice for the Implementation of Renewable Energy Forecasting Solutions Part Three: Forecast solution selection} \citep{recomprac}. The package is tailored for evaluation of ensemble forecasts, which can both be evaluated as probabilistic forecasts using appropriate metrics, or as event forecasts using traditional classifier evaluation tools. Until now, there has been no tool available capable of handling both forms of evaluation of ensembles, especially with respect to event detection. Helping to close this gap is a main contribution of \Rpack.\\

The primary motivation for the release of the package is the abundant use of ensembles in forecasting of weather and renewables. The popularity of ensembles is attributed to the forecast uncertainty of wind speeds and solar radiation, which is costly for wind and solar power producers, traders and consumers \citep{sorensen2023recent}. Ensembles provide a way to cover the variety of possible weather scenarios such that large financial losses can be mitigated and profits maximized. The renewable energy industry is responsible for an increasing share of the power mix in Europe. For instance, the share of renewables increased to 47\% in Q3 of 2024, up from 43\% in Q3 of 2023, while the share of fossil fuels went down in the same period \citep{q3}. With higher renewable penetration, the need for proper evaluation of probabilistic forecasts intensifies.\\

Some existing R packages for forecast evaluation include \texttt{s2dverification} \citep{manubens2018r} for evaluation of climate forecasts, \texttt{ForecastTB} \citep{bokde2020forecasttb} for evaluation of forecasts using traditional metrics such as RMSE and MAE, and \texttt{metrica} \citep{correndo2022metrica} for evaluation of point forecasts.\\

The methods provided by \Rpack\text{ }include the continuous ranked probability score (CRPS), the logarithmic score (LogS), the variogram score (VarS), the Brier score, the transformed rank histogram, the reliability diagram, the reciever operating characteristic (ROC) curve and the contingency table and are briefly described in Section 2 and 3 with their original references. Furthermore, the package offers a user-friendly framework for applying the methods to forecast and observation data, as long as the data is organized as required (see Section \ref{sec:organize}).\\
\section{Probabilistic forecast evaluation}\label{sec:probscores}

Probabilistic forecasts are forecasts that, in contrast to point forecasts, contain some information about the forecast uncertainty. Examples are full forecast densities, a set of quantiles or a simple confidence interval. In \Rpack\text{ }, ensemble forecasts are treated as representative samples from true stochastic processes. The users are themselves responsible for ensuring that this assumption is upheld for the ensembles they evaluate with the package.

\subsection{Scoring rules}

Let $F(x)$ be the cumulative distribution function(CDF) of a forecast and let $y$ be the observation that materalizes. The CRPS \citep{matheson1976scoring} evaluates the full marginal distribution and is defined as follows:

\begin{equation}
    \text{CRPS}(F,y) = \int_{x \in \mathbb{R}} (F(x)-\mathbb{I}(y > x))^2 dx.
\end{equation}

This score is known for being robust with respect to highly unlikely events. While a multivariate version called the energy score exists, it has been shown to lack the ability to detect misspecified correlation well and is hence not considered here.\\

As an alternative, the log score (LogS)  \citep{gneiting2007strictly} evaluates the probability density function (PDF) $f(x)$ and is defined as:

\begin{equation}
    \text{LogS}(f,y) = -\log(f(y)).
\end{equation}

The LogS evaluates the entire multivariate distribution with a superb ability to seperate models with different correlation structures. However for numerical computation the PDF has to be estimated which is expensive or even infeasible for higher dimension. Thus, in practice this metric is best suited for univariate problems. In contrast to the CRPS, it is also very sensitive to unlikely observations. For some applications, extreme penalties on the unlikely can be a desirable property \citep{bjerregaard2021introduction}.\\

For multivariate problems, the variogram score (VarS) of order $p$ \citep{scheuerer2015variogram} has a strong ability to detect misspecified correlation structure and is computationally feasible for high-dimensional problems. It is defined as follows:

\begin{equation}
    \text{VarS}(\bm{F},y) = 
    \sum_{i=1}^{n-1} \sum_{j=i+1}^{n} w_{ij} (|y_i - y_j|^p - E(X_i - X_j))^2,
\end{equation}

where $\bm{F}$ is the multivariate (joint) forecast distribution, $X_i$ is the $i$'th marginal forecast distribution of $\bm{F}$ and $p$ is usually set to 0.5. The VarS is a sum of score contributions from pairs of marginal forecast distributions and $w_{ij}$ is the weight given to the contribution by the pair $X_i$ and $X_j$. Some typical weight choices are $w_{ij} = 1$ or $w_{ij} = 1/|i-j|$.\\

The main drawback of the VarS is that it cannot separate uncalibrated from calibrated forecasts because it is invariant under changes to the mean. Therefore, the user should generally ensure that forecasts are calibrated before they are evaluated by the VarS.

\subsection{Rank histogram}

In addition to scoring rules, it is useful to visually assess how well forecasts are calibrated. This can be accomplished by constructing a rank histogram.\\

In the traditional rank histogram, for each marginal ensemble forecast, the ensemble members together with the observation are sorted and assigned ranks according to their position in the sorted list. For example, an ensemble of $m = 20$ members will have 21 ranks, and if an observation falls right in the middle of the list, it will get rank 11.\\

\begin{figure}[H]
    \centering
    \includegraphics[width=\textwidth]{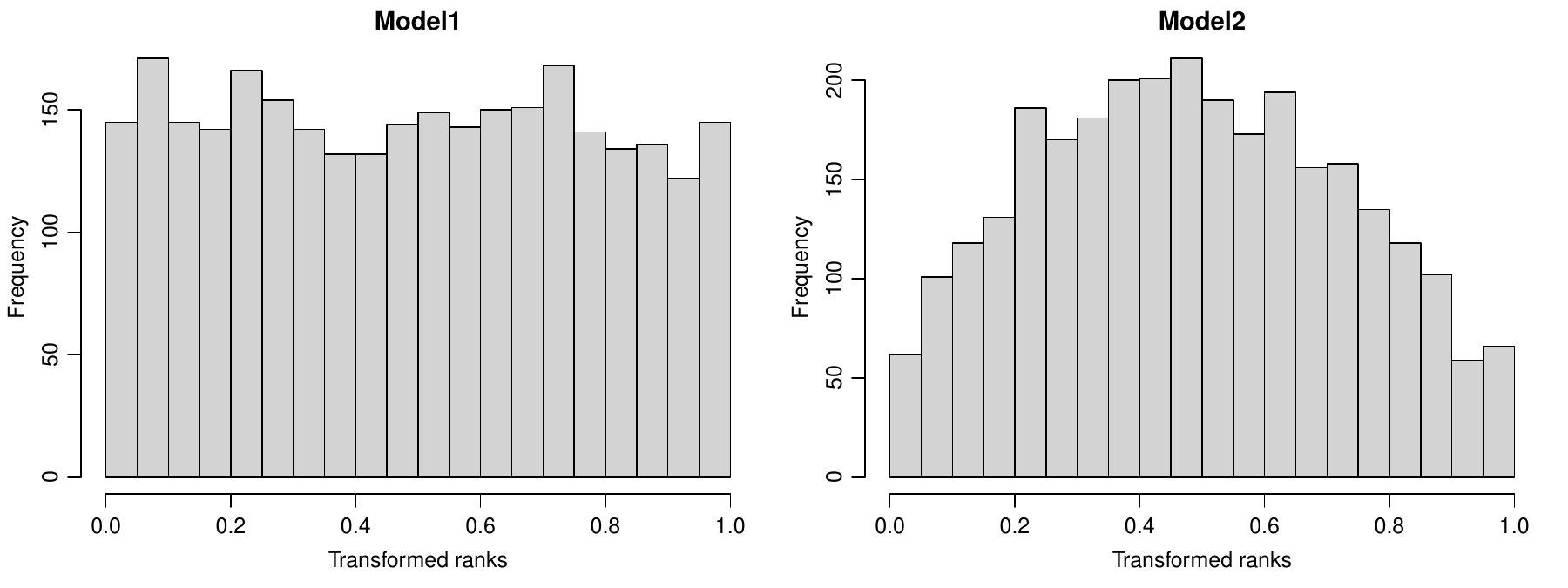}
    \vspace{-0.7cm}
    \caption{Rank histograms of the two forecast models in \datB, with 3-hour lead time.}
    \label{fig:rankhist}
\end{figure}

One issue with the traditional rank histogram is that its representativeness relies on the chosen number of bins. For example, if 10 bins are chosen to make a rank histogram from an ensemble with $m = 20$ (21 ranks), one of the bins will hold one more rank than the others, making it harder to assess the calibration of the forecast. This problem is addressed in \Rpack\text{ } by using the transformed rank histogram \citep{heinrich2021number} which transforms the rank $i$ in the following way:

\begin{equation}
    \text{rank}_{i,\text{transformed}} = \frac{\text{rank}_i - 1 + U_i}{m + 1},
\end{equation}

where $m$ is the number of ensemble members and $U_i \sim U(0,1)$ is a uniformly distributed variable, independent for all ranks. This means, the rank histogram will change every time it is recomputed. A seed can be set in order to get fully reproducible results. Some examples of transformed rank histograms are seen in Fig. \ref{fig:rankhist}.
\section{Event-based forecast evaluation}\label{sec:event}

For some applications, it is crucial to forecast the timing and magnitude of certain events with a high degree of precision. For example, rapid changes in wind power production have a huge impact on the electricity price and getting the timing, sign or magnitude wrong can lead to considerable financial losses for power traders.

\Rpack\text{ }supports event-based forecast evaluation on different levels. State-of-the-art evaluation methods available in the package follow the recommendations of \cite{messner2020evaluation} and \cite{recomprac} and include Brier scores, receiver operating characteristic (ROC) curves, reliability diagrams and contingency tables. Furthermore, the \texttt{event\_detection\_table} function provides a flexible way to search a forecast dataset for custom-defined events and return a list of all event detections, which can then be evaluated using the aforementioned methods.

\subsection{Event characterization}

\begin{figure}[H]
    \centering
    \includegraphics[width=\textwidth]{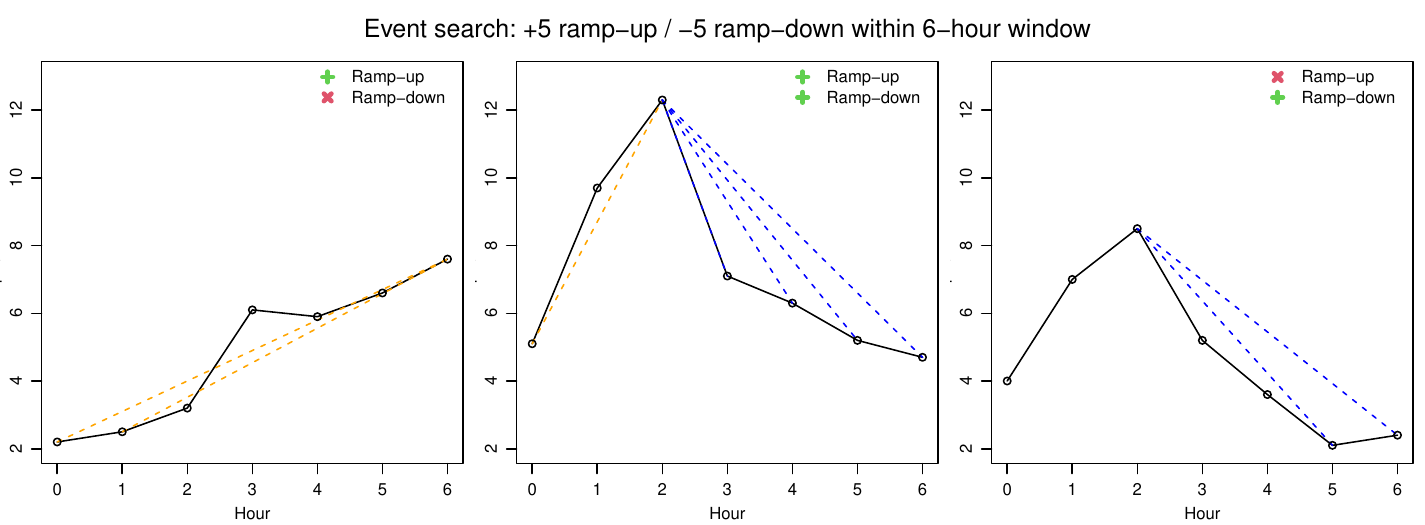}
    \vspace{-0.7cm}
    \caption{Examples of event search.}
    \label{fig:events}
\end{figure}

There are multiple ways to define events of interest, and there is no consensus of one correct way to do it or one complete list of every thinkable event. Here, we consider two main types of events which cover most situations. For example, in the context of forecasting of renewables, these would be the most likely critical event types for grid operators, traders or renewables portfolio managers. For both main types, a rolling time window is defined within which the event is searched for:
\begin{itemize}
    \item A \textbf{range-event} occurs when the variable of interest attains a value within a defined range or interval at least once during the searched time window. For example, we can assess the ability to detect periods with wind speeds over 12 m/s by considering the interval (12, $\infty$).
    \item A \textbf{change-event} occurs when the variable of interest undergoes a signed change anywhere during the searched time window, including for data points that are not next to each other (see Fig. \ref{fig:events}). Positive/negative changes are denoted ramp-up and ramp-down, respectively. For example we can consider a flat increase in normalized power production by 0.3, i.e. a ramp of 30\% of installed capacity.
\end{itemize}
The concept of ramp-up and ramp-down is illustrated in Fig. \ref{fig:events}.

\begin{figure}[H]
    \centering
    \includegraphics[width=\textwidth]{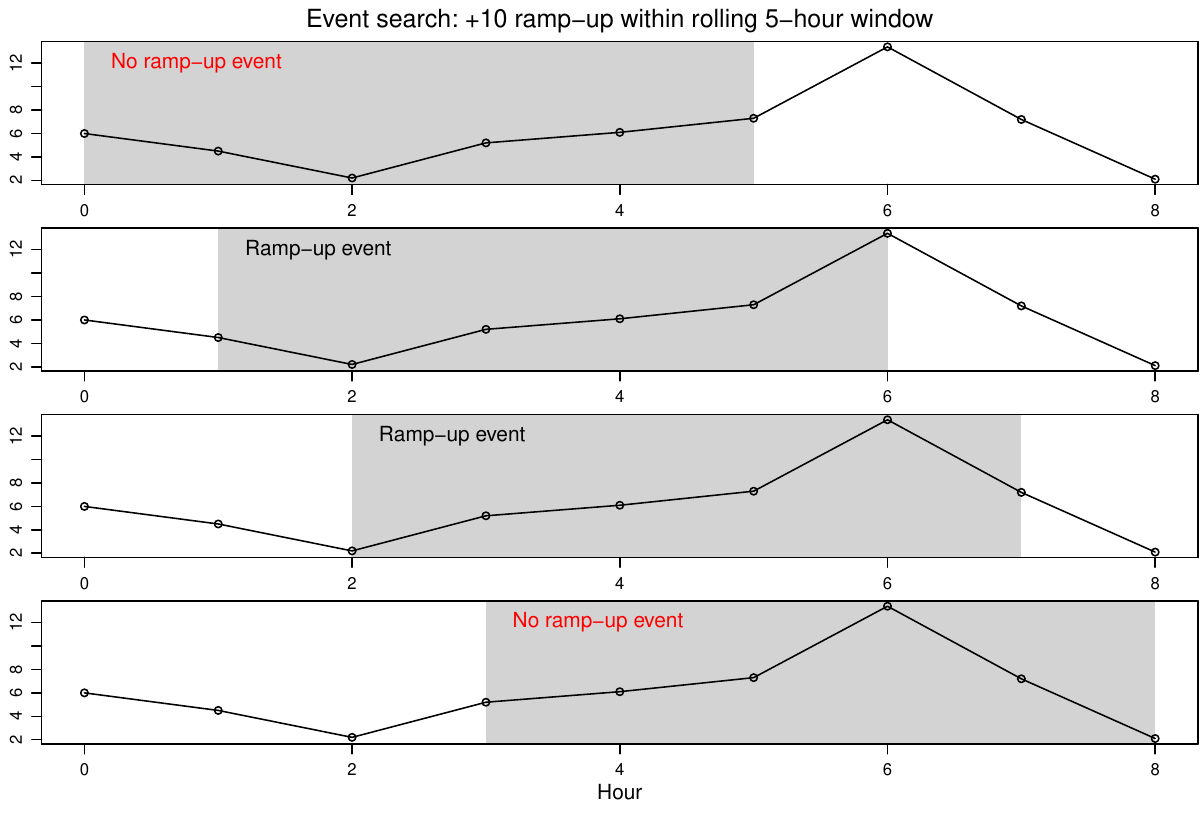}
    \vspace{-0.7cm}
    \caption{Example of a rolling window of event searches.}
    \label{fig:rollingwindow}
\end{figure}

When searching a full forecast dataset for events, a rolling time window is considered, which means the same event can happen multiple times if the window is wider than the time resolution, as illustrated in Fig. \ref{fig:rollingwindow}.

\subsection{Brier score}

The Brier score is similar to the RMSE, but is specifically tailored for evaluation of event forecasts.
Let $o = (o_1, \cdots, o_{N_e})$ be a list of $N_e$ observed event outcomes, where

\begin{equation}
    o_i =
    \begin{cases}
        1 & \text{if event in window $i$}\\
        0 & \text{if no event in window $i$}
    \end{cases}.
\end{equation}

Given the list of corresponding forecast probabilities $f = (f_1, \cdots, f_{N_e})$, where every $f_i$ is a fraction defined on the interval $[0,1]$, the Brier score is defined as:

\begin{equation}
    \text{BS}(f,o) = \frac{1}{N_e} \sum_{i=1}^{N_e} (f_i - o_i)^2.
\end{equation}

A low Brier score indicates a high accuracy on event forecasting. This metric is often separated into the three components: reliability, resolution and uncertainty \citep{murphy1973new}, i.e.:
\begin{equation}
    \text{BS}(f,o) = REL - RES + UNC
\end{equation}
with
\begin{equation}
    REL = \frac{1}{N_e}\sum_{k=1}^K n_k (f_k - \bar{o}_k)^2 ,\quad RES = \frac{1}{N_e}\sum_{k=1}^K n_k (\bar{o}_k-\bar{o})^2,\quad UNC = \bar{o}(1-\bar{o})
\end{equation}
where $K$ is the number of different forecast probabilities, $n_k$ is the number of forecasts with the $k$'th forecast probability ($f_k$), $\bar{o}_k$ is the observed frequency of events within the subset with forecast probability $f_k$ and finally, $\bar{o} = \frac{1}{N_e} \sum_{i=1}^{N_e} o_i$ is the overall observed frequency of events.

\subsection{Reliability diagram}

The reliability diagram provides insight to how well the forecasted probabilities of events ocurring match the observed relative frequencies. For example, given a forecast dataset, if there are 1000 events which are each forecasted to be 70\% chance of occring, then 700 of them should turn out as events, and 300 of them as non-events. The same is true for every probability between 0\% and 100\%. In practice, finite binning is used to evaluate probability intervals rather than single probabilities. A typical binning consists of the following 11 bins:

\begin{center}
    \begin{tabular}{c|c}
    Bin no. & Probability interval \\
    \hline
    1 & $0-5\%$ \\
    2 & $5-15\%$ \\
    3 & $15-25\%$ \\
    4 & $25-35\%$ \\
    5 & $35-45\%$ \\
    6 & $45-55\%$ \\
    7 & $55-65\%$ \\
    8 & $65-75\%$ \\
    9 & $75-85\%$ \\
    10 & $85-95\%$ \\
    11 & $95-100\%$ \\
\end{tabular}
\end{center}

After defining the bins, the forecasts are grouped according to their predicted probabilities, and the observed relative frequencies is then calculated for each bin as the number of events divided by the total number of event forecasts in the same bin:
\begin{equation}
    \text{obs. relative frequency}_k = \text{obs. occurences}_k / \text{num. forecasts}_k.
\end{equation}
For example, if there are 45 event forecasts in a given bin, and the observed outcomes amounts to 17 events and 28 non-events, then the observed relative frequency of that bin is 17/45 = 0.3777.
Calculation of the forecast probabilities depends on the model. For ensemble forecast, the simplest way is to consider the forecast probability as the fraction of ensemble members that has predicted an event to occur. Hence, if 5 members out of an ensemble of 50 predicts an event to happen, then the forecast probability is 5/50 = 0.1.

\begin{figure}[H]
    \centering
    \includegraphics[width=0.7\textwidth]{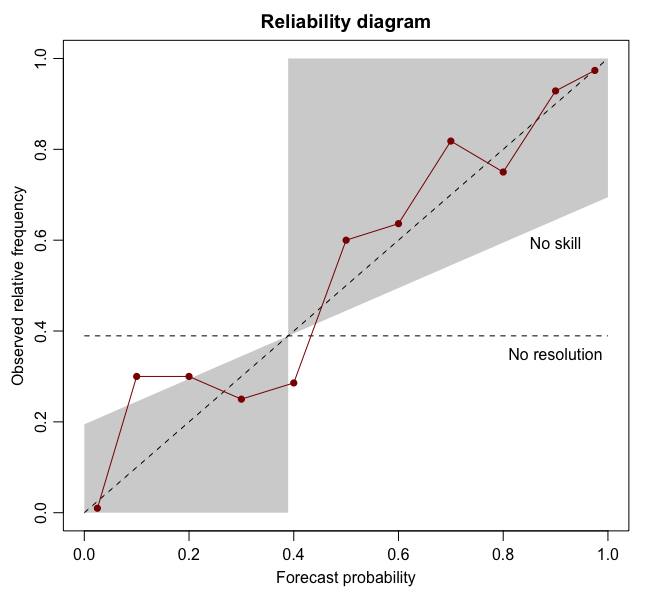}
    \caption{A reliability diagram.}
    \label{fig:reldiag1}
\end{figure}

When all pairs of forecast proabilities and observed relative frequencies have been calculated, the reliability diagram can be drawn as exemplified in Fig. \ref{fig:reldiag1}. The diagonal line connecting (0,0) with (1,1) corresponds to perfect reliability.

\subsection{ROC curve and contingency table}

The ROC (reciever operating characteristic) curve highlights the ability of a forecast to classify binary events correctly, without causing too many false alarms. It is a 2-dimensional curve connecting pairs of true positive rates and false alarm rates.

In order to construct the curve, we consider the four possible outcomes of a binary event forecast which altogether form a \textbf{contingency table}:

\begin{center}
    \begin{tabular}{l|cc}
    & Event happened & No event happened \\
    \hline
    Predicted event & True positive (TP) & False positive (FP) \\
    Predicted no event & False negative (FN) & True negative (TN) \\
\end{tabular}
\end{center}

Given a forecast dataset of $N_e$ event outcomes, we obtain the \textbf{true positive rate} (TPR):
\begin{equation}
    \text{TPR} = \frac{\sum \text{TP}}{\sum \text{TP} + \sum \text{FN}}
\end{equation}
and the \textbf{false alarm rate} (FAR):
\begin{equation}
    \text{FAR} = \frac{\sum \text{FP}}{\sum \text{TP} + \sum \text{FP}}
\end{equation}

\begin{figure}[H]
    \centering
    \includegraphics[width=0.7\textwidth]{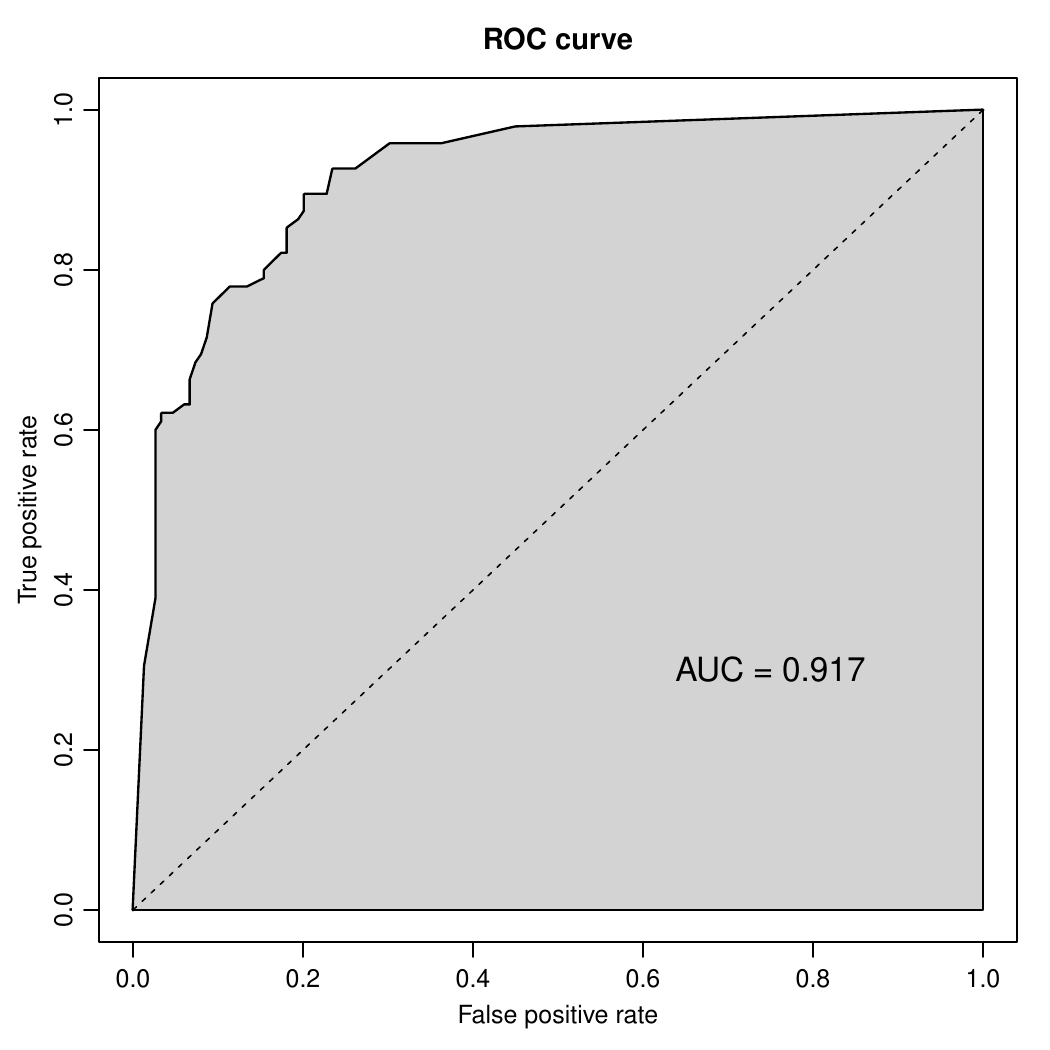}
    \caption{A ROC curve.}
    \label{fig:roc1}
\end{figure}

Calculating the TPR and FAR for a forecast dataset only gives one data point. In order to get the full ROC curve, an event detection criterion must be selected and varied. In the case of ensemble forecasts, a typical criterion is that a certain number of ensemble members must predict the event to happen in order for the final forecast to be an event. 
By varying the threshold, a list of (FAR,TPR)-pairs are obtained that can be plotted as the ROC curve, see Fig. \ref{fig:roc1}. A lower threshold on required ensemble members leads to more event detections but at the cost of more false alarms, while a higher threshold reduces the number of false alarms at the cost of fewer successful event detections. A skilled forecast model should produce a high true positive rate with a simultaneously low false positive rate.

Furthermore, as an overall measure of the event detection skill, the area under the curve (AUC) is computed and is equal to 1 for a forecast which is always correct and 0.5 (diagonal line) for random guessing.

\section{Program structure}\label{sec:organize}
\subsection{Forecasting and observation data}

The package works on forecasting and observation data which is organized in a specific way. The data structure is a list of two elements, one being the forecasts and the other one being the observations. The forecast element is itself a list of each individual forecasting candidate. Each forecasting candidate or observation series is a data frame. There is currently only support for univariate observation series. The hierarchical structure of the data structure is illustrated below:\\

\begin{tikzpicture}
\draw (0,-0.75) -- (0,-0.25) node[anchor=south] {data structure};
\draw (0,-0.75) -- (3,-0.75) -- (3,-1) node[anchor=north] {observations};
\draw (0,-0.75) -- (-3,-0.75) -- (-3,-1) node[anchor=north] {forecasts};
\draw (-3,-1.6) -- (-3,-2);
\draw (3,-1.6) -- (3,-3.75);
\draw (-5,-2) -- (-1,-2);
\draw (-5,-2) -- (-5,-2.5) node[anchor=north] {forecast1};
\draw (-3,-2) -- (-3,-2.5) node[anchor=north] {forecast2};
\draw (-1,-2) -- (-1,-2.5) node[anchor=north] {forecast3};
\draw[dashed] (-1,-2) -- (-0.25,-2);
\draw (3,-2) -- (3,-3.75) node[anchor=north] {obs};
\draw (-5,-3.2) -- (-5,-3.5);
\draw (-6,-3.5) -- (-4,-3.5);

\draw (-6,-3.5) -- (-6,-3.75) node[anchor=north] {m1};
\draw (-5,-3.5) -- (-5,-3.75) node[anchor=north] {m2};
\draw (-4,-3.5) -- (-4,-3.75) node[anchor=north] {m3};
\draw[dashed] (-4,-3.5) -- (-3.5,-3.5);

\draw[dashed] (-3,-3.1) -- (-3,-3.4);
\draw[dashed] (-1,-3.1) -- (-1,-3.4);

\end{tikzpicture}
\vspace{0.4cm}\\
Furthermore, the data frames must comply with a certain formatting. For each forecast data frame:
\begin{itemize}
    \item The first column must be named "\texttt{TimeStamp}" (case-sensitive) and be in POSIXct format.
    \item An optional column with timestamps at which the forecasts were issued must be named "\texttt{BaseTime}", if it exists. It must be in POSIXct format as well.
    \item There are no requirements on the labeling of the forecast member columns, and the format of the values must be numeric.
\end{itemize}

For the observation data frame:
\begin{itemize}
    \item The first column must be named "\texttt{TimeStamp}" (case-sensitive) and be in POSIXct format.
    \item The second column contains the observations. It must be named "\texttt{obs}" and be in numeric format.
\end{itemize}

After the data structure has been created, it can then be saved to an rda-file for later use. \Rpack\text{ }also allows the user to store all forecast and observation data as csv-files and run the \texttt{load\_forecast\_data} function to build the appropriate data structure.\\

Note that the methods in \Rpack\text{ }are suited for ensemble forecasts with multiple members, so evaluation of one-member (essentially deterministic) forecasts should be addressed by other tools.

\subsection{Function usage}

\begin{figure}[H]
    \centering
    \includegraphics[width=\textwidth]{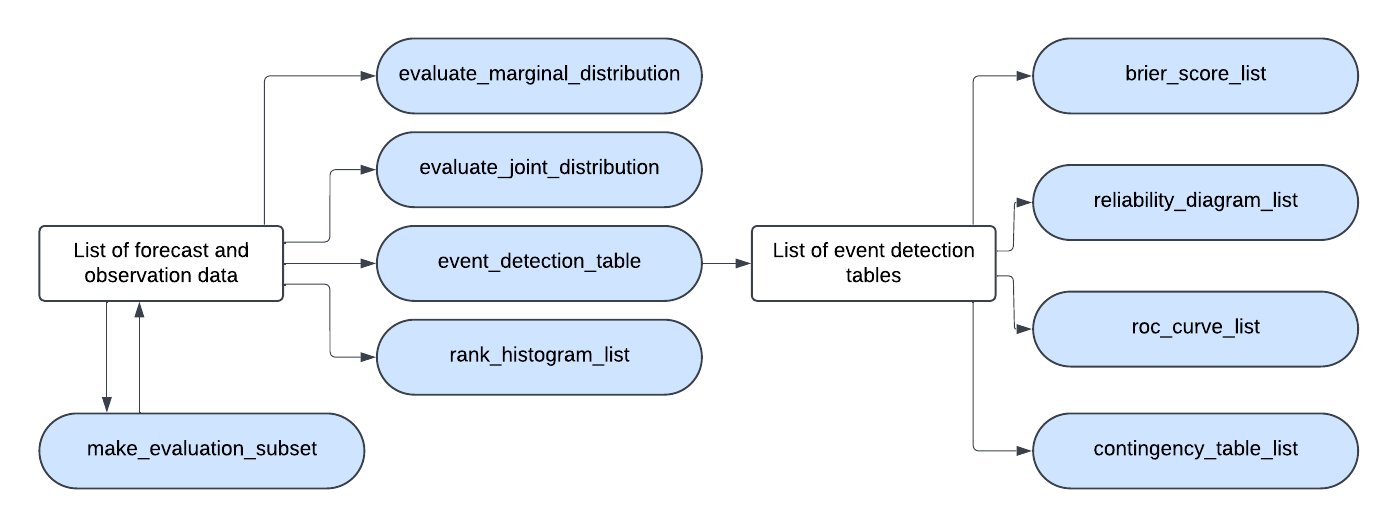}
    \vspace{-0.7cm}
    \caption{The path from data structure to forecast evaluation. Data is shown as white nodes and \Rpack{ } functions are shown as light blue nodes.}
    \label{fig:usage}
\end{figure}

When the data structure defined above is followed, forecast evaluation can be performed by calling high-level functions designed to run over all models and all forecasts within each model and apply the desired evaluation method. For the event-based evaluation methods, an input of event detections must be supplied. It is possible to construct this input from raw forecasts using the function \texttt{event\_detection\_table}, but the user is also free to generate it by external means. The process from data to evaluation application is visualized in Fig. \ref{fig:usage}.

All of these high level functions call lower level functions which are freely accessible to the user. For example, if the user has prepared a table of event probabilities and observations and want to construct a reliability diagram from it, then this can be accomplished by using the \texttt{reliability\_diagram} function directly rather than \texttt{reliability\_diagram\_list}.
\section{Practical examples}

In this section, practical use of \Rpack\text{ } is demonstrated in two examples each demonstrating different features. Some previous examples can be found in \cite{mohrlen2023iea}.

\subsection{Example 1: Univariate and multivariate probabilistic forecast evaluation on 14 days of ensemble forecasts}

This example uses the simulated \datA\text{ }dataset, which consists of 14 days of hourly wind speed observations, as well as forecasts issued every 3 hours by two competing models labeled Model1 and Model2.\\

Assuming the data is located in the current working directory, it can be imported with \texttt{load}:

\begin{lstlisting}[frame=single, style=R]  % Start your code-block

load("simplewind.rda")

\end{lstlisting}

As a start, the content of the data can be checked with \texttt{summary\_stats} which displays the first 6 lines of the observations and forecasts, as well as useful metrics such as the mean, minimum, maximum, number of data points and number of missing values. The purpose of the function is to quickly verify that the dataset is imported correctly and that it does not contain obvious errors.\\

\begin{figure}[H]
    \centering
    \includegraphics[width=\textwidth]{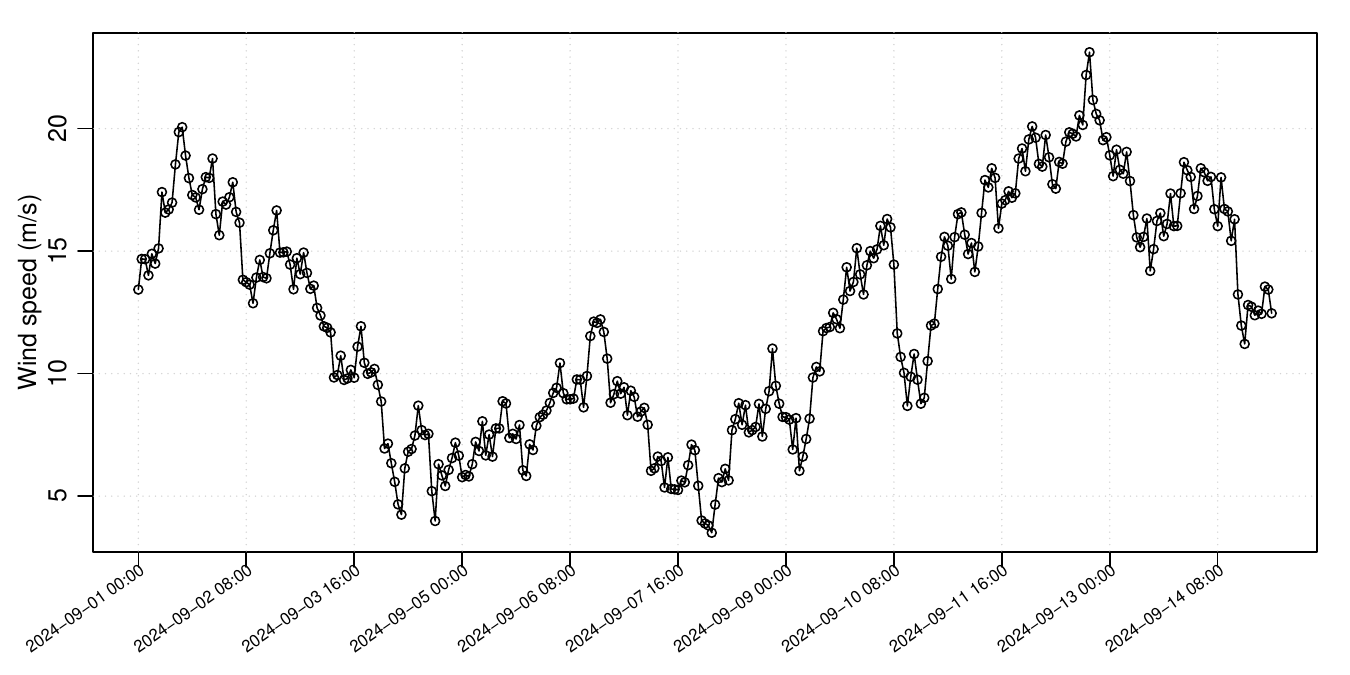}
    \vspace{-0.7cm}
    \caption{All observations from the \datA\text{ }dataset.}
    \label{fig:evex_obs}
\end{figure}

Furthermore, the observations can be plotted with generic plot functions or the \texttt{plot\_observations} included in \Rpack. Example code is as follows:
\newpage

\begin{lstlisting}[frame=single, style=R]  % Start your code-block

data <- simplewind

plot_observations(data$observations, all = T, grid = T)

summary_stats(data)

\end{lstlisting}

The observations as visualized by \texttt{plot\_observations} are shown in Fig. \ref{fig:evex_obs}. Here, the argument \texttt{all=T} has been used to show all observations in the dataset. If the dataset is very large, it can be convenient to leave out this argument, in which case only the first 100 observations are shown. Alternatively, the argument \texttt{numobs} can be used to show a specific number of observations.

The output from \texttt{summary\_stats} is as follows:

\begin{lstlisting}[style=output]
OBSERVATIONS
------------
           TimeStamp   obs
 2024-09-01 00:00:00 13.43
 2024-09-01 01:00:00 14.63
 2024-09-01 02:00:00 14.63
 2024-09-01 03:00:00 13.99
 2024-09-01 04:00:00 14.83
 2024-09-01 05:00:00 14.45


FORECASTS
---------
Model1
           TimeStamp   BaseTime    m1    m2    m3
 2024-09-01 01:00:00 2024-09-01 11.42 10.60 14.41
 2024-09-01 02:00:00 2024-09-01 11.46 12.05 14.89
 2024-09-01 03:00:00 2024-09-01 12.75 12.22 14.36
 2024-09-01 04:00:00 2024-09-01 11.51  9.93 12.74
 2024-09-01 05:00:00 2024-09-01 13.76 11.75 13.82
 2024-09-01 06:00:00 2024-09-01 18.94 10.75 14.25

Model2
           TimeStamp   BaseTime    m1    m2    m3
 2024-09-02 01:00:00 2024-09-02 18.03 17.56 13.27
 2024-09-02 02:00:00 2024-09-02 15.60 16.60 14.53
 2024-09-02 03:00:00 2024-09-02 18.84 16.44 17.81
 2024-09-02 04:00:00 2024-09-02 19.27 18.02 18.48
 2024-09-02 05:00:00 2024-09-02 21.58 20.04 17.33
 2024-09-02 06:00:00 2024-09-02 19.70 18.58 17.04


SUMMARY
-------
$observations
     mean  min   max number_of_observations missing_values
 12.32077 3.95 22.76                    337              0

$Model1
    mean  min   max number_of_forecasts ensemble_size
 12.1342 0.07 28.52                5424            20

$Model2
     mean  min   max number_of_forecasts ensemble_size
 12.07727 0.04 36.17                4632            20
\end{lstlisting}


Like for the observations, \Rpack\text{ } offers a way to plot the forecast data with the \texttt{plot\_forecasts} function. The forecasts must be ensembles for it to work, because the plot is based on quantiles which are computed by the function in an internal step.\\

Forecast data usually contains forecasts with different lead times and sometimes overlapping forecasts issued at different base times can be dealt with by setting the optional argument \texttt{by} to either \texttt{"leadtime"} or \texttt{"basetime"}. If \texttt{by="leadtime"} is used, the function extracts all of the forecasts with one specific lead time, defaulting to 1 hour. The specific lead time can be set with the \texttt{lead} argument in units of hours. If \texttt{by="basetime"} is used, all forecasts issued at the first base time of the dataset are extracted. If \texttt{by} is not set, then any forecasts with overlapping timestamps are filtered out before plotting.\\

\begin{figure}[H]
    \centering
    \includegraphics[width=\textwidth]{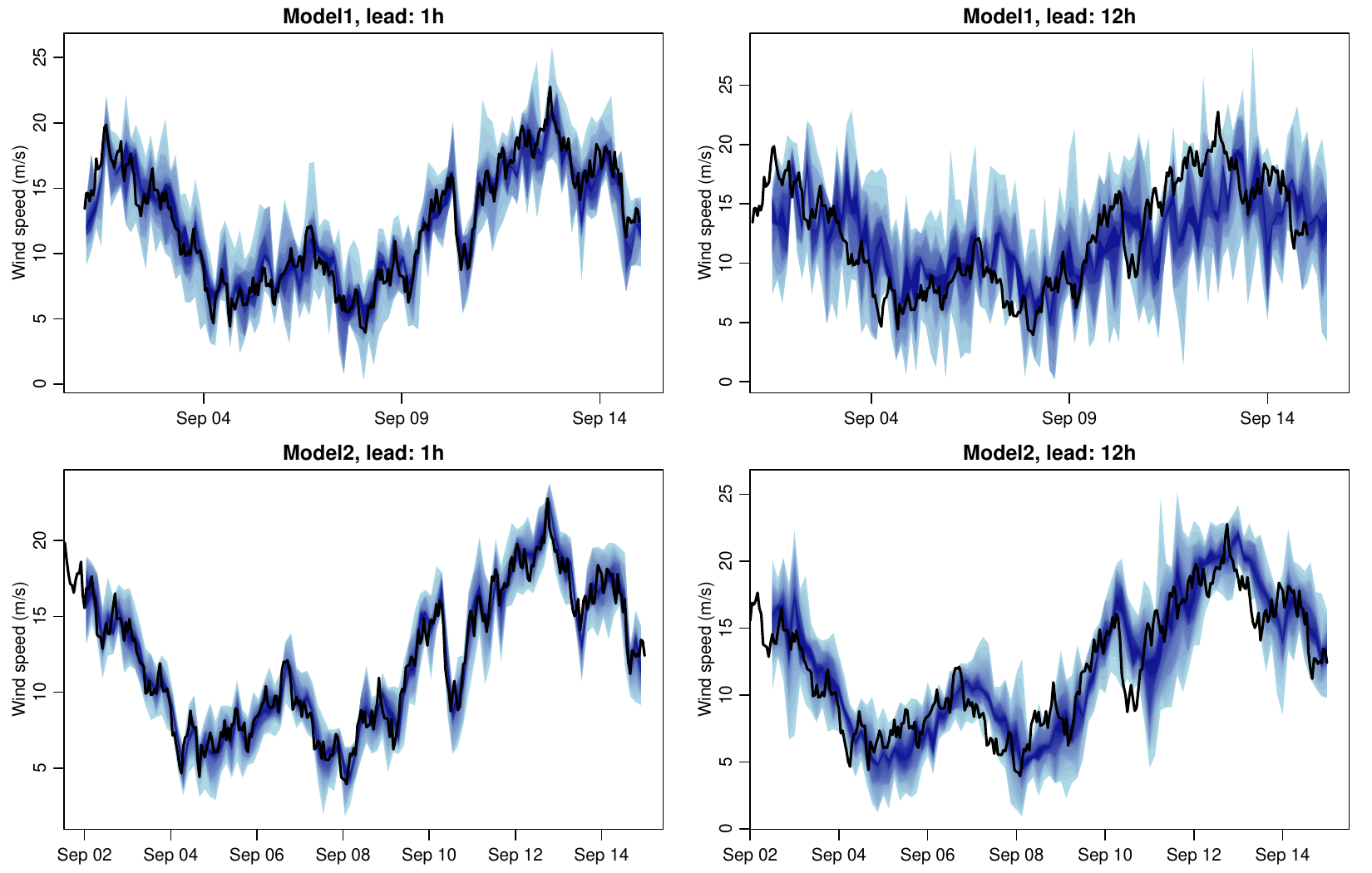}
    \vspace{-0.7cm}
    \caption{Forecasts from the \datA\text{}dataset issued by 2 different models both with lead times of 1 hour and 12 hours.}
    \label{fig:evex_forec}
\end{figure}

Example code could be the following:

\begin{lstlisting}[frame=single, style=R]  % Start your code-block

plot_forecasts(data$forecasts$Model1, by = "leadtime", lead = 1, all = T)
lines(data$observations, lwd=2)

\end{lstlisting}

In Fig. \ref{fig:evex_forec} the code is repeated for both forecast models with \texttt{lead} set to 1 or 12 hours. The figure gives the impression that Model2 is superior and that both models perform better at the 1-hour horizon than at the 12-hour horizon.

When the user is confident that both the observation and forecast data look reasonable, a number of evaluation methods can be applied depending on the problem at hand. In this example, the goal is to evaluate the forecast densities produced by the two models, both as univariate and multivariate densities. The univariate part is performed by the \texttt{evaluate\_marginal\_distribution} function. This function evaluates every marginal ensemble forecast with its associated observation, one timestamp at a time. Timestamps that do not have both an observation and a forecast are automatically filtered out. Two different metrics can be selected, namely the CRPS (default) or the LogS (see Section \ref{sec:probscores}). As for the plotting, the presence of different lead times in the dataset can be handled by setting the argument \texttt{by\_lead\_time=T}.\\

Example code with CRPS as well as truncated output are shown below:

\begin{lstlisting}[frame=single, style=R]  % Start your code-block

scores_crps <- evaluate_marginal_distribution(data, by_lead_time = T, 
                                              metric = "CRPS")
plot_score_by_leadtime(scores_crps)

\end{lstlisting}

\begin{lstlisting}[style=output]
    forecast leadtime      CRPS
1     Model1        1 0.7807045
2     Model1        2 1.1025884
3     Model1        3 1.3861232
...............................
46    Model1       46 2.7345072
47    Model1       47 2.7766206
48    Model1       48 2.8658577
49    Model2        1 0.5870279
50    Model2        2 0.7752663
51    Model2        3 0.8562707
...............................
94    Model2       46 1.2702935
95    Model2       47 1.2582337
96    Model2       48 1.3184725
97 reference       NA 2.5632579
\end{lstlisting}

\begin{figure}[H]
    \centering
    \includegraphics[width=\textwidth]{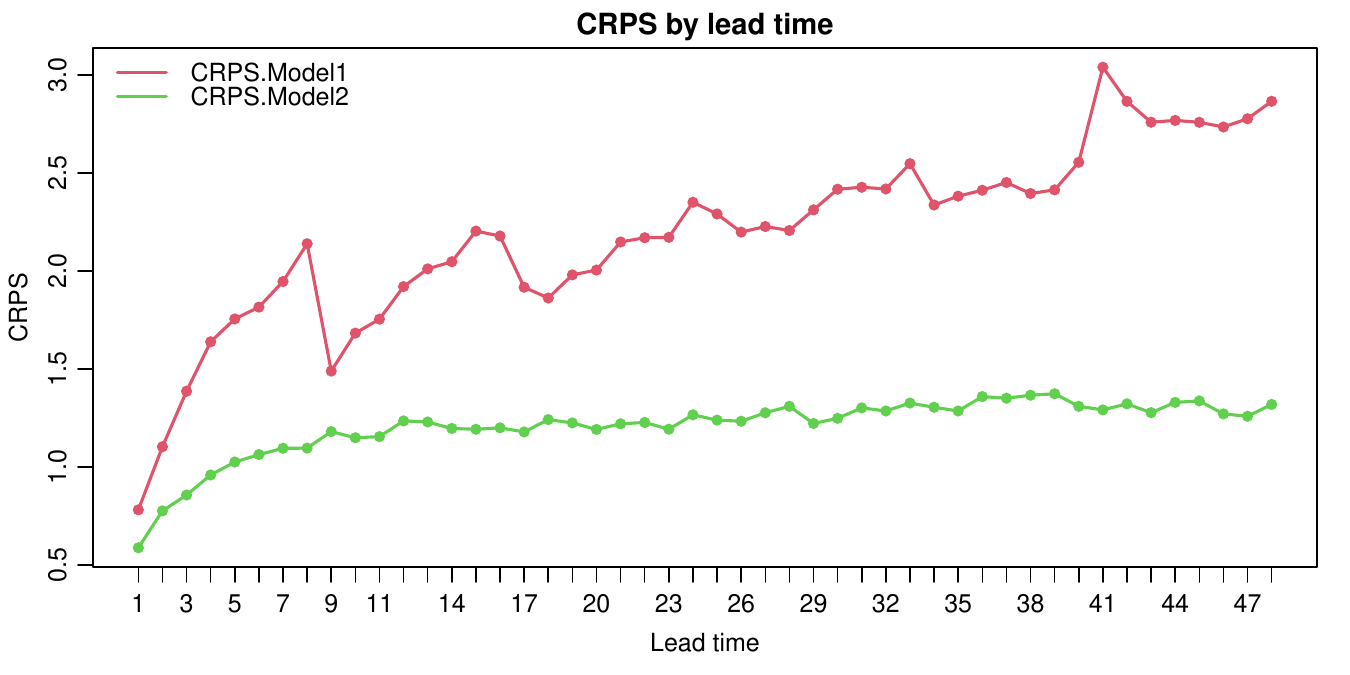}
    \vspace{-0.7cm}
    \caption{Evaluation of the forecast ensembles in \datA\text{} separated by lead time.}
    \label{fig:evex_crps}
\end{figure}

The output from \texttt{evaluate\_marginal\_distribution} can be plotted conveniently by the specialized \texttt{plot\_score\_by\_leadtime} function, with the result shown in Fig. \ref{fig:evex_crps}.

For the multivariate part, the function \texttt{evaluate\_joint\_distribution} is used. Hereby, all marginal ensemble forecasts issued at the same base time by the same model are regarded as a multivariate ensemble forecast with temporal dimension equal to the number of marginal forecasts. For instance, in the \datA\text{ }dataset, the forecasts are 1-48 hour ahead, so most of them are 48-dimensional except for some forecasts close to the ends of the data period. Similarly to the marginal case, the code is simply:

\begin{lstlisting}[frame=single, style=R]  % Start your code-block

scores_vars <- evaluate_joint_distribution(data, by_base_time = T)

\end{lstlisting}

The output is a variogram score for each multivariate forecast, i.e. one per model per base time, which is shown in truncated form below (the full output has 226 lines):

\begin{lstlisting}[style=output]
    forecast            basetime        VarS dimension
1     Model1 2024-09-01 00:00:00  85.1490976        48
2     Model1 2024-09-01 03:00:00 112.4924597        48
3     Model1 2024-09-01 06:00:00 106.4262134        48
......................................................
95    Model1 2024-09-12 18:00:00 130.8261607        48
96    Model1 2024-09-12 21:00:00 117.3509359        48
97    Model1 2024-09-13 00:00:00 103.2267473        48
98    Model1 2024-09-13 03:00:00 137.7854432        45
99    Model1 2024-09-13 06:00:00  95.2197097        42
100   Model1 2024-09-13 09:00:00  59.8235893        39
......................................................
111   Model1 2024-09-14 18:00:00   4.9983559         6
112   Model1 2024-09-14 21:00:00   0.8183297         3
113   Model1 2024-09-15 00:00:00          NA        NA
114   Model2 2024-09-01 00:00:00          NA        NA
115   Model2 2024-09-01 03:00:00          NA        NA
......................................................
122   Model2 2024-09-02 00:00:00  54.3800517        48
123   Model2 2024-09-02 03:00:00  48.3880910        48
124   Model2 2024-09-02 06:00:00  50.6901390        48
......................................................
223   Model2 2024-09-14 15:00:00   6.8843467         9
224   Model2 2024-09-14 18:00:00   5.3687693         6
225   Model2 2024-09-14 21:00:00   0.5900366         3
226   Model2 2024-09-15 00:00:00          NA        NA
\end{lstlisting}

\begin{figure}[H]
    \centering
    \includegraphics[width=\textwidth]{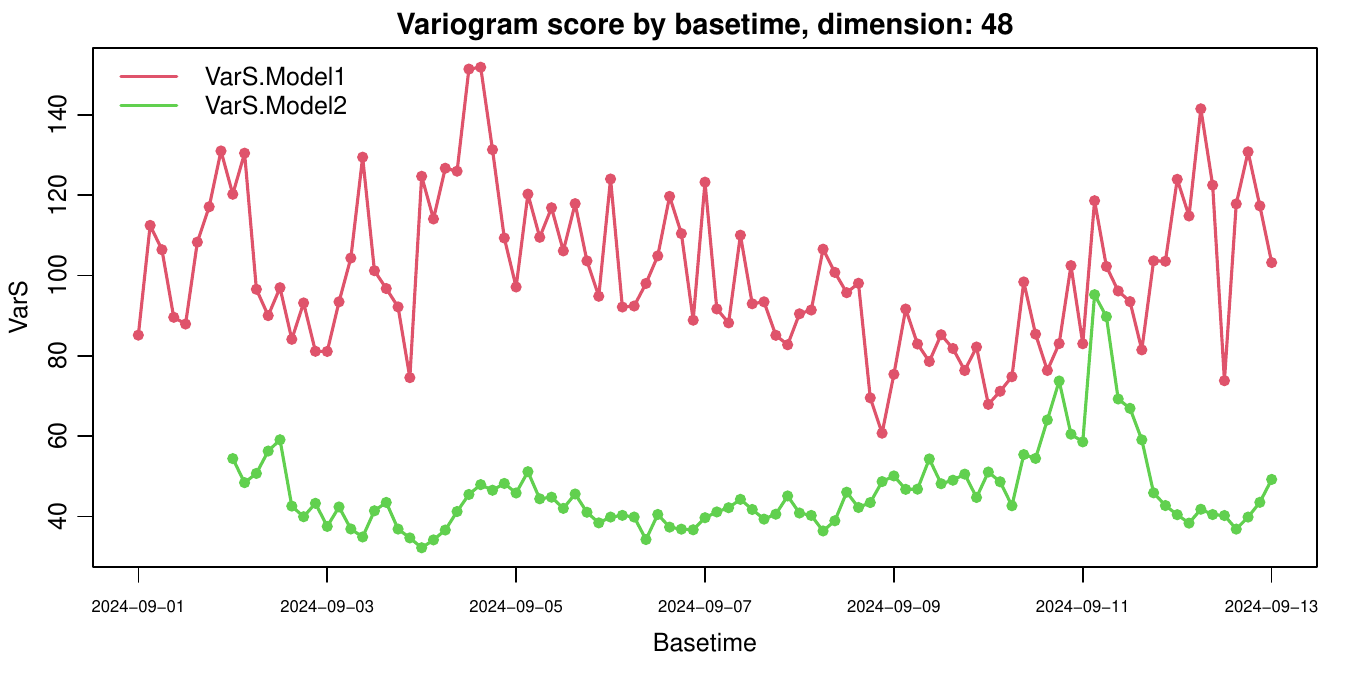}
    \vspace{-0.7cm}
    \caption{Evaluation of the forecast ensembles in \datA\text{} as multivariate forecasts issued for every base time.}
    \label{fig:exvars}
\end{figure}

The average VarS for each model can be computed by setting the optional argument \texttt{aggregate=T}. As multivariate forecasts with different dimension are not meaningful to compare, setting this argument will automatically throw away all of the multivariate forecasts that do not have the maximal dimension. Hence, the code

\begin{lstlisting}[frame=single, style=R]  % Start your code-block

evaluate_joint_distribution(data, by_base_time = T, aggregate = T)

\end{lstlisting}

gives the output:

\begin{lstlisting}[style=output]
  forecast      VarS dimension
1   Model1 100.60701        48
2   Model2  46.07793        48
\end{lstlisting}

The non-aggregated output can be visualized with the function \texttt{plot\_variogram\_scores}, which will likewise only use the multivariate forecasts with maximal dimension. Both the plot (Fig. \ref{fig:exvars}) and the overall scores clearly selects Model2 as the best at capturing the temporal correlation of the wind speed process.

Thus, it has been demonstrated how a time series of ensembles may be evaluated as both univariate and multivariate probalistic forecasts with \Rpack.
\subsection{Example 2: Event detection on one year of ensemble forecasts}

This example is based on a \datB\text{}, a dataset similar to that in Example 1. Again it features wind speed observations which are forecasted by the same two competing models, however this time there is a full year of data. The example highlights how to filter the data for a certain lead time, how to perform event detection and evaluation as well as how to make rank histograms.\\

\begin{figure}[H]
    \centering
    \includegraphics[width=\textwidth]{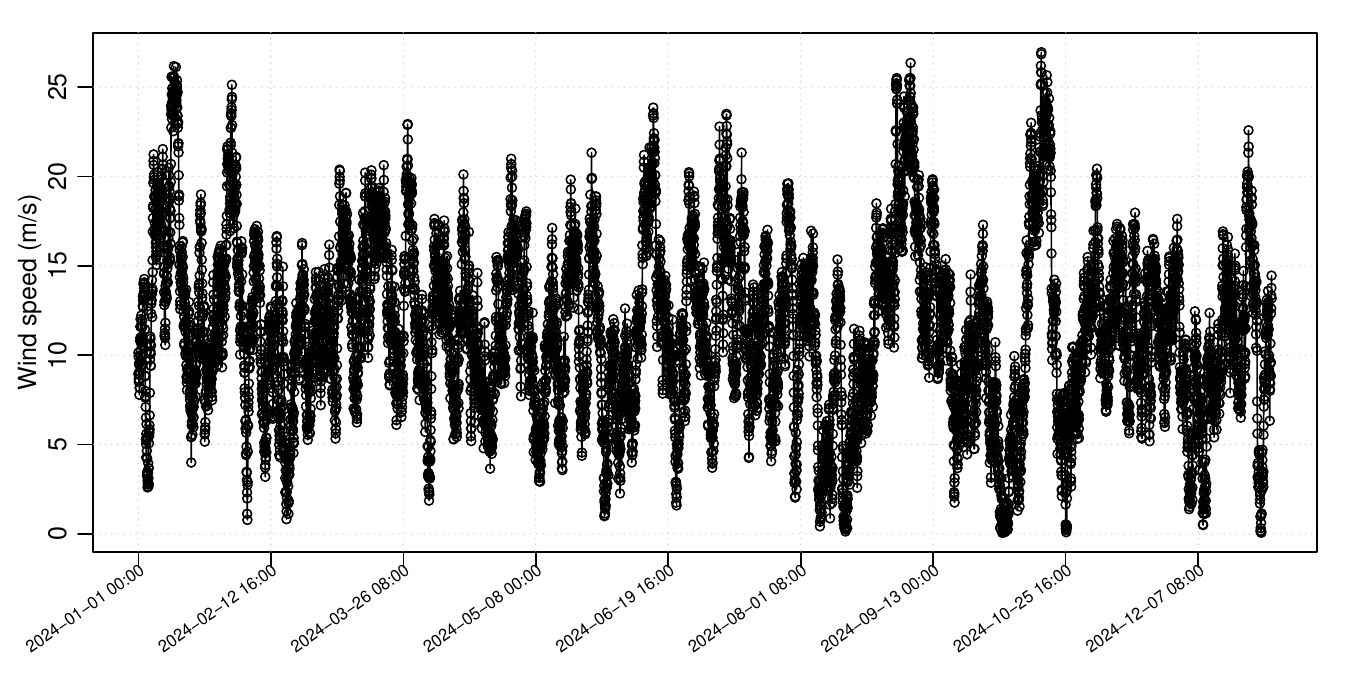}
    \vspace{-0.7cm}
    \caption{All observations from the \datB\text{ }dataset.}
    \label{fig:evex2_obs}
\end{figure}

The observations are shown in Fig. \ref{fig:evex2_obs}, the forecasts are not shown here, but they are very similar to those in Fig. \ref{fig:evex_forec}.

As in Example 1, the dataset contains 48 different lead times. It can be filtered for a specific lead time by calling the \texttt{make\_evaluation\_subset} function and set the \texttt{lead\_time} argument. The resulting subset can then be subject to forecast evaluation. For instance, rank histograms can be produced by the functions \texttt{rank\_histogram} and \texttt{rank\_histogram\_list}, where the former takes a forecast data frame and an observation data frame as arguments while the latter acts on the full data structure and returns rank histograms for every forecast model hereof.\\

Below is some example code where a lead time of 3 hours is chosen and followingly evaluated with rank histograms. It is seen that Model1 appears to be well calibrated, while Model2 is overdispersive.

\begin{lstlisting}[frame=single, style=R]  % Start your code-block

data <- oneyearwind
dat_eval <- make_evaluation_subset(data, lead_time = 3)
rank_histogram_list(dat_eval)

\end{lstlisting}

\subsubsection*{Detection and evaluation of a range-event}

In the rest of the example, we focus on event detection and evaluation on forecasts with 24-hour lead time. The event detection is performed by the \texttt{event\_detection\_table} function which acts on the usual forecast-observation data structure and returns a table for each forecast model with the results. The tables have the same format as the original data frames, but with 1 and 0 entries for events and non-events, respectively, instead of the original numerical (in this case wind speed) values. Then, the event detections can be evaluated with ROC curves, reliability diagrams, brier scores or contingency tables, see Section \ref{sec:event}.\\

Two different types of events are tested. We define the first event as the presence of at least one entry above 12 m/s within a 3-hour window. Example code is as follows:

\begin{lstlisting}[frame=single, style=R]  % Start your code-block

dat_eval <- make_evaluation_subset(data, lead_time = 24)
events <- event_detection_table(dat_eval, range = c(12,Inf), window = 3)

roc_curve_list(events)

reliability_diagram_list(events)

brier_score_list(events)

contingency_table_list(events, threshold = 0.5)

\end{lstlisting}

The beginning of the output from \texttt{event\_detection\_table} function for Model1 looks like the following:

\begin{lstlisting}[frame=single, style=output]
            TimeStamp            BaseTime obs m1 m2 m3 m4 m5 m6 ... m18 m19 m20
1 2024-01-03 00:00:00 2024-01-02 00:00:00   1  0  1  0  1  1  0 ...   0   1   0
2 2024-01-03 03:00:00 2024-01-02 03:00:00   0  0  0  0  0  1  0 ...   0   1   0 
3 2024-01-03 06:00:00 2024-01-02 06:00:00   0  1  0  0  0  1  1 ...   0   1   1
4 2024-01-03 09:00:00 2024-01-02 09:00:00   0  1  1  1  0  1  1 ...   0   0   1
  .............................................................................

\end{lstlisting}

\begin{figure}[H]
    \centering
    \includegraphics[width=\textwidth]{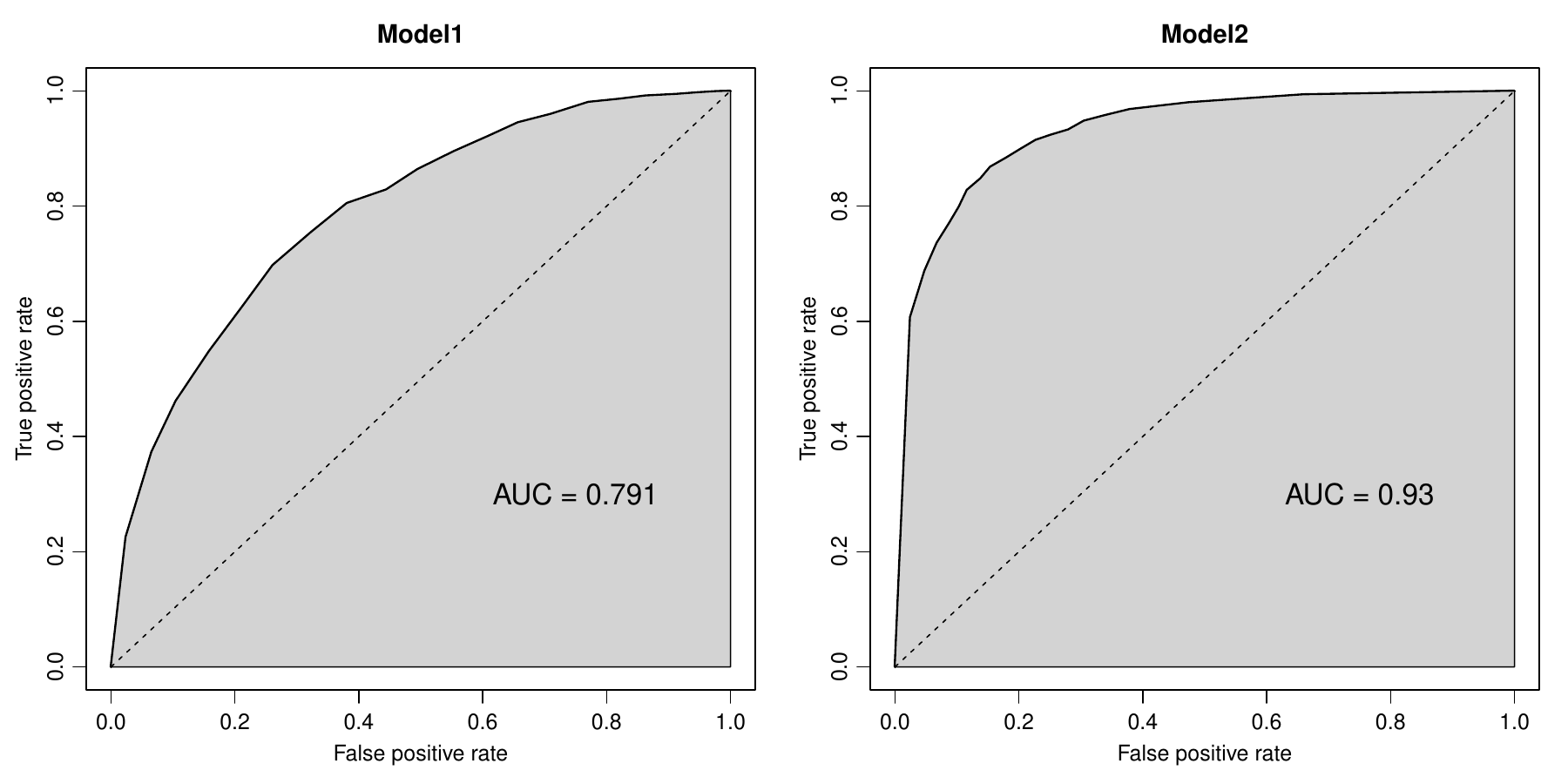}
    \includegraphics[width=\textwidth]{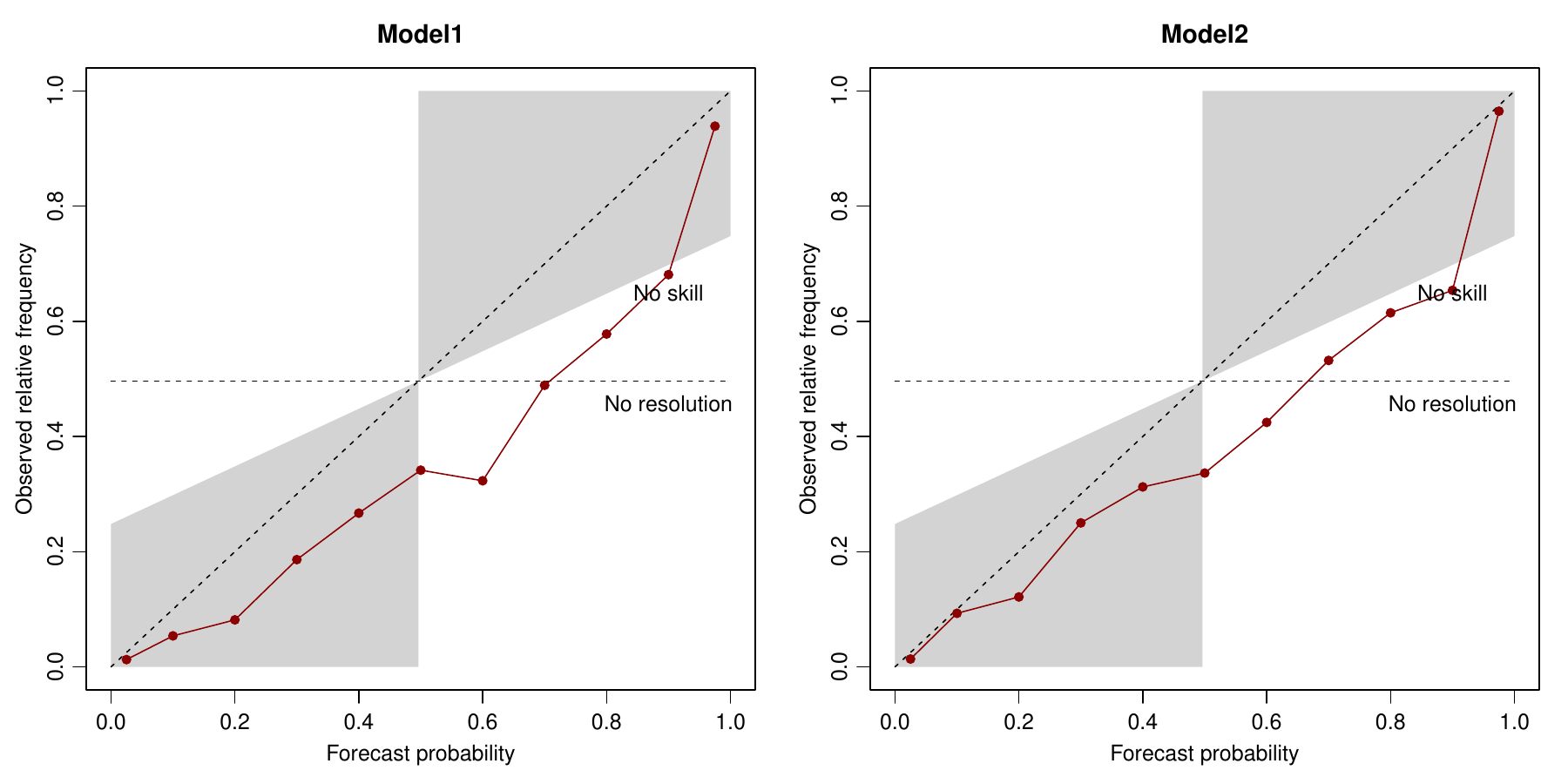}
    \vspace{-0.7cm}
    \caption{ROC curves (top row) and reliability diagrams (bottom row) of the two forecasting models, based on range-event detections.}
    \label{fig:ev1_roc}
\end{figure}

The ROC curves and the reliability diagrams of both forecast models are shown in Fig. \ref{fig:ev1_roc}. It appears that Model2 is better at detection this particular event than Model1 according to the ROC curve, while it is difficult to see any obvious difference in terms of reliability.

The Brier scores report Model2 to be the best as well, as 0.101 is lower than 0.194:

\begin{lstlisting}[frame=single, style=output]
   Model1    Model2 
0.1940009 0.1012623 
\end{lstlisting}

Finally, the contingency table shows that Model 2 is capable of producing a 91,5\% hit rate while keeping the false alarm rate at 22,7\%. The threshold was set to 0.5 in the code, which means 10 out of the 20 members must return 1 to consider it an event. Recall, this HR/FAR-pair is a point on the ROC curve in Fig. \ref{fig:ev1_roc}, where all the other points are obtained by setting different thresholds. The contingency table for Model 1 reports worse numbers, which is consistent with the corresponding ROC curve:

\begin{lstlisting}[frame=single, style=output]
$Model1
  hits misses falsealarms correctnegatives    HR   FAR
1 1290    151         810              653 0.895 0.554

$Model2
  hits misses falsealarms correctnegatives    HR   FAR
1 1318    123         332             1131 0.915 0.227
\end{lstlisting}

\subsubsection*{Detection and evaluation of a change-event}

The second event detection example follows the same recipe as the previous, but with the event defined as a 0.5 decrease in wind speed within a 3-hour window. The code and output is shown below:

\begin{lstlisting}[frame=single, style=R]  % Start your code-block

events <- event_detection_table(dat_eval, change = -0.5, window = 3)

\end{lstlisting}

\begin{figure}[H]
    \centering
    \includegraphics[width=\textwidth]{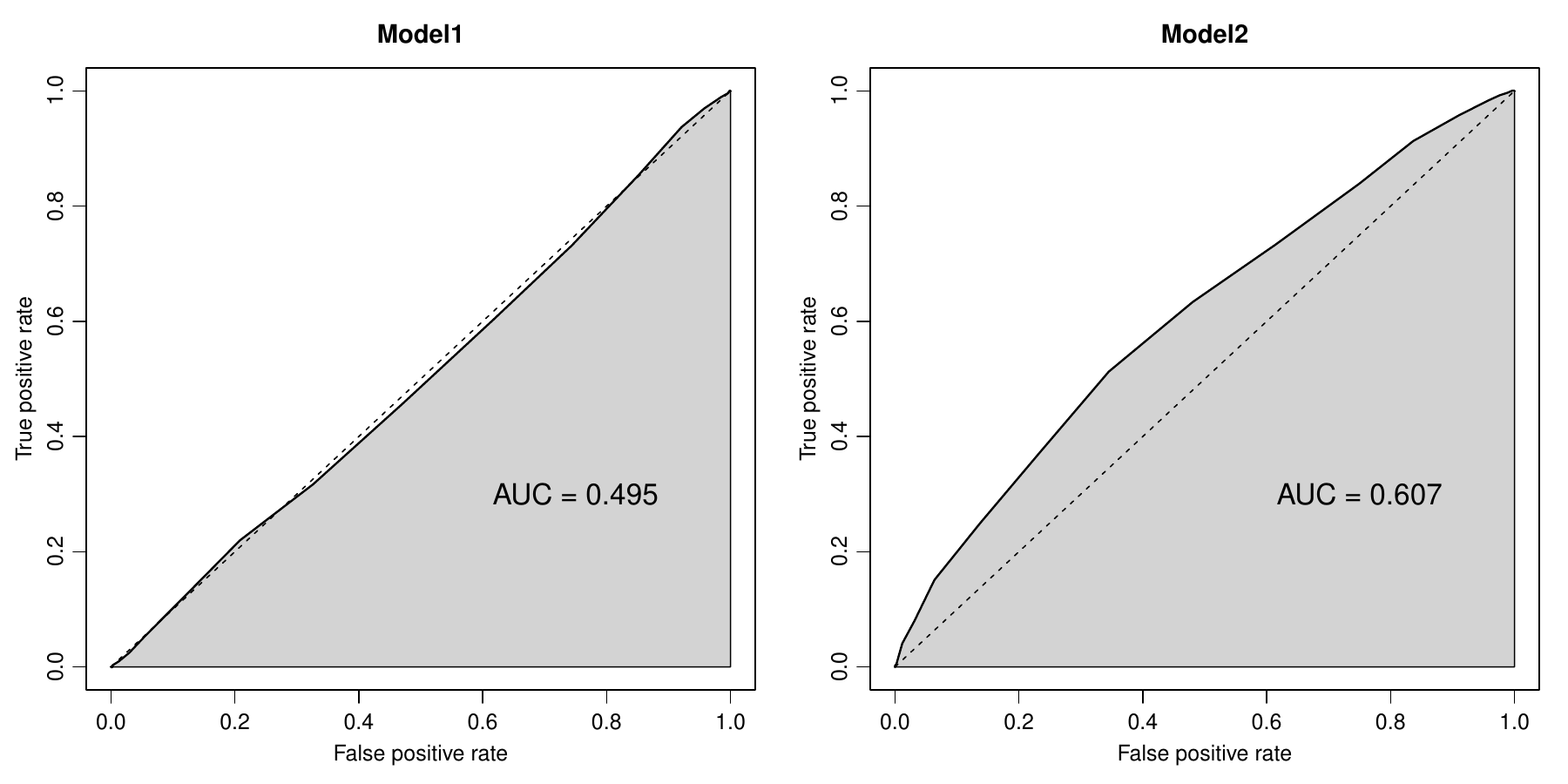}
    \includegraphics[width=\textwidth]{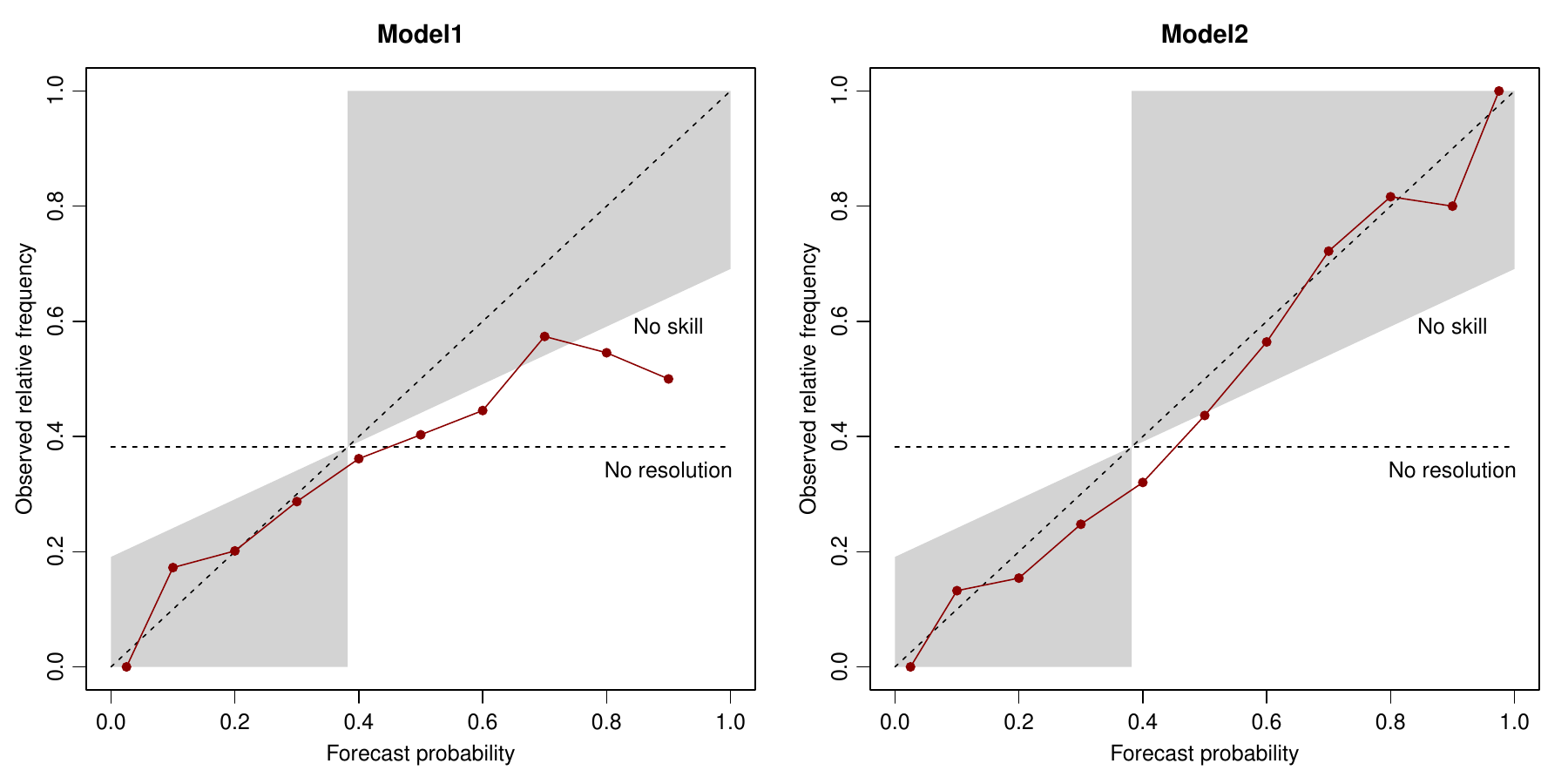}
    \vspace{-0.7cm}
    \caption{ROC curves (top row) and reliability diagrams (bottom row) of the two forecasting models, based on change-event detections.}
    \label{fig:ev2_roc}
\end{figure}

Brier score output:

\begin{lstlisting}[frame=single, style=output]
   Model1    Model2 
0.2371418 0.2124913 
\end{lstlisting}

\newpage

Contingency tables:

\begin{lstlisting}[frame=single, style=output]
$Model1
  hits misses falsealarms correctnegatives    HR   FAR
1 1039     70        1654              141 0.937 0.921

$Model2
  hits misses falsealarms correctnegatives    HR   FAR
1 1012     97        1502              293 0.913 0.837
\end{lstlisting}

Interestingly, the reliability diagram of Model2 looks better for the change-events than it did for the range-events, while the ROC curve and the contingency table show that it is in fact not very selective and cannot detect the events without producing a lot of false alarms. Model1 once again performs more poorly than Model2 in all of the evaluations. Especially the ROC curve reveals that random guessing would be slightly better at detecting this particular event than Model1.
\section{Conclusion}

The new R-package \Rpack\text{ }offers a flexible tool for evaluation of forecast ensembles, both for the users who are interested in evaluating the densities as well as those interested in the ability to forecast events. With a well-defined structure to organize the forecast and observation data, state-of-the art probabilistic forecast evaluation can be done by \Rpack\text{ }in very few lines of code. However, as the underlying evaluation methods are implemented in separate lower level functions, more advanced users have the option to build their own evaluation setups hereon. The development of \Rpack\text{ }is expected to continue in the future, and voluntary contribution from the community is welcome.

\section{Acknowledgements}

The authors gratefully acknowledge IEA Wind Task 51 (EUDP project
134-22015), Weather2X (EUDP project 640232-511303) and DynFlex (part of
the Danish Mission Green Fuel projects).

\bibliographystyle{plainnat}
\bibliography{cites}

\end{document}